\documentclass[a4paper,12pt]{article}
\usepackage{amsmath}
\usepackage{amsfonts}
\usepackage{mathtools}
\usepackage{authblk}
\usepackage{graphicx,epsfig}
\usepackage{float}
\usepackage{subcaption}
\usepackage{amssymb}
\usepackage{xcolor}
\usepackage{cite}

\numberwithin{equation}{section}
\title{Black hole shadow  and acceleration bounds for spherically symmetric spacetimes}

\author[1,2]{Kajol Paithankar\footnote{kajol.paithankar@iiap.res.in}\,}
\author[1]{Sanved Kolekar\footnote{sanved.kolekar@iiap.res.in} \,}
\affil[1]{Indian Institute of Astrophysics, Koramangala II Block, Bangalore 560034, India}
\affil[2]{Pondicherry University, R.V. Nagar, Kalapet 605014, Puducherry, India}

\begin{document}

\maketitle

\begin{abstract}

We explore an interesting connection between black hole shadow parameters and the acceleration bounds for radial linear uniformly accelerated (LUA) trajectories in static spherically symmetric black hole spacetime geometries of the Schwarzschild type. For an incoming radial LUA trajectory to escape back to infinity, there exists a bound on its magnitude of acceleration and the distance of closest approach from the event horizon of the black hole. We calculate these bounds and the shadow parameters, namely the photon sphere radius and the shadow radius, explicitly for specific black hole solutions in $d$-dimensional Einstein's theory of gravity, in pure Lovelock theory of gravity and in the $\mathcal{F}(R)$ theory of gravity. We find that for a particular boundary data, the photon sphere radius $r_{ph}$ is equal to the bound on radius of closest approach $r_b$ of the incoming radial LUA trajectory while the shadow radius $r_{sh}$ is equal to the inverse magnitude of the acceleration bound $|a|_b$ for the LUA trajectory to turn back to infinity. Using the effective potential technique, we further show that the same relations are valid in any theory of gravity for static spherically symmetric black hole geometries of the Schwarzschild type. Investigating the trajectories in a more general class of static spherically symmetric black hole spacetimes, we find that the two relations are valid separately for two different choices of boundary data.

\end{abstract}

\section{Introduction}

One of the most interesting features of a black hole is the shadow seen in its image. The idea of observing the black hole shadow using Very Long Baseline Interferometry (VLBI) goes way back to the start of this century and was first proposed by Falcke et al. in \cite{Falcke} wherein an image of Sagittarius A* was simulated using general relativistic ray-tracing code. It was proposed that with the resolution of global VLBI arrays at about 1.3 mm wavelength, the shadow of Sgr A* should be observable. Later, images of two black holes, M87* and Sgr A* were successfully produced by the Event Horizon Telescope (EHT) collaboration \cite{EHT 2019,EHT 2022}, which provided the first ever direct visual evidence of these black holes and has motivated a detailed study of various aspects of black hole shadows.

The first simulated image of a Schwarzschild black hole surrounded by a thin accretion disk was constructed by J. P. Luminate where the boundary of the black hole shadow was defined by the marginally trapped light rays \cite{Luminet}. A comprehensive study of null trajectories and the marginally trapped light rays has been done for the four dimensional black holes in Einstein's gravity, namely the Schwarzschild black hole \cite{Synge,Zeldovich}, the Reissner–Nordström black hole \cite{Zakharov 1,Zakharov 2} and the Kerr black hole \cite{Bardeen,Viergutz}. Some of the other black hole solutions for which the shadow size has been evaluated are black holes of Pleba\'nski-Demia\'nski class, a black hole surrounded by dark matter, a class of regular black holes, braneworld black holes etc. A detailed review of analytical studies of these black hole shadows can be found in \cite{Review} and references therein. Such investigations of the shadow are not only restricted to the black holes in Einstein's gravity but are also extended to alternative and modified theories of gravity such as, the Lovelock gravity \cite{Lovelock gravity}, the scalar tensor theory\cite{scalar tensor gravity}, the $\mathcal{F}(R)$ gravity \cite{f(R) gravity} etc. Many more black hole solutions such as the charged, rotating black holes in the Einstein Gauss Bonnet gravity, $\mathcal{F}(R)$ gravity, $f(T)$ gravity and Rastall gravity have also been investigated for their shadows, see \cite{ Lovelock 1, Lovelock 2, Lovelock 3, fR 1, fR 2, fT, Rastall 1, Rastall 2, Rastall 3, Virbhadra, 4D EGB, Expanding universe, Randall-Sundrum AdS5 brane-world, NLED, scalar hair, Many different models} for examples. 

An interesting relationship has been found between the quasi-normal modes (QNMs) derived in the eikonal limit and the shadow radius of static spherically symmetric asymptotically flat  black holes. The real part of the QNM is related to the angular velocity of the circular photon orbit while the imaginary part of the QNM is connected to the Lyapunov exponent which is related to the stability or instability of the orbits. The relationship between the QNMs and the shadow of a rotating black hole has been checked for a variety of cases including the rotating black hole solutions in Einstein's gravity (see \cite{Cardoso, Jusufi, Yang, Konoplya Stuchlik, Konoplya}) as well as static spherically symmetric black hole solutions in theories of gravity beyond Einstein's gravity (see \cite{QNM1,QNM2,QNM3,QNM4,QNM5,QNM6,QNM7,QNM8,QNM9}). 

The investigation of black hole shadows essentially involves the study of null geodesics around the curved black hole background. On the other hand, one could study accelerated trajectories in a black hole geometry, particularly the linear uniformly accelerated (LUA) trajectories. The LUA trajectories are curved spacetime generalization of linear uniformly accelerated Rindler trajectories in flat spacetime \cite{Rindler}. The LUA curves satisfy the Letaw-Frenet equations for fixed curvature scalar and vanishing torsion and hyper-torsion \cite{Letaw,Kolekar}. In the usual flat spacetime picture, a Rindler trajectory with constant magnitude of acceleration $|a|$ lying in the $(X-T)$ plane and confined to, say, the right Rindler wedge, formed by the past and future horizon null surfaces $X=-T$ and $X=T$ respectively, where $X$ and $T$ are Minkowski coordinates is a LUA trajectory. The casual structure of Rindler quadrant along with the time-like boost Killing trajectories leads to the well known Unruh effect \cite{Davies,Unruh}. An interesting aspect is to understand how the quadrant structure and the globally hyperbolic motion of the Rindler trajectories are modified due to spacetime curvature when a gravitating body is brought in the picture. An extreme case, would be to study the LUA trajectories in the background of a static black hole spacetime. Since an event horizon already exists in the black hole geometry, a Rindler type motion, in particular the LUA motion, would introduce additional Rindler-type horizons in the co-moving frame of the LUA observer. Interesting questions arise in such a scenario in a semi-classical setting as to what does the LUA observer see in the Hartle-Hawking vacuum or in the Unruh vacuum. For the Schwarzschild spacetime, radial LUA trajectories were investigated in \cite{Kajol} and the corresponding Rindler quadrant structure was analysed in \cite{Sanved}. Interestingly, it was found that there exist an upper bound on the magnitude of acceleration of the radial LUA trajectories incoming from past infinity to turn back at some radius; else the trajectory falls into the black hole horizon. This acceleration bound further leads to a bound on the radius of turning point, that is, the distance of closest approach for these LUA trajectories to escape to infinity. It is these parameters, the acceleration bound and the turning point bound of the LUA trajectories that we investigate and compare with the black hole shadow parameters of the unstable circular photon orbit.  

In this work, we investigate and report a connection between the black hole shadow parameters and the acceleration bounds on radial linear uniformly accelerated (LUA) trajectories for static spherically symmetric black holes of the Schwarzschild type. The paper is organized as follows. In \ref{The Shadow}, we first set up the basic null geodesic equations in a Schwarzschild type spacetime and summarise the definitions related to the shadow parameters in section \ref{Basics of shadow}. We then evaluate the shadow parameters of static spherically symmetric black holes in three different theories of gravity, namely, the Einstein's theory, the Lovelock theory and the $\mathcal{F}(R)$ theory. In particular, we evaluate the shadow radii of the Einstein's charged black hole in $d$ dimensions with $d\geq 4$, vacuum black hole solution in pure Lovelock theory and a black hole with conformally coupled scalar field in $\mathcal{F}(R)$ theory of gravity in sections \ref{Shadow of RN}, \ref{Shadow of Pure LL} and \ref{Shadow of f(R)} respectively. The shadows of black holes in Einstein's theory have been well studied in four dimensions. Consideration of higher dimensions then accounts for some correction to the size of the shadow calculated for a four dimensional black hole. The pure Lovelock theory have many interesting features. The vacuum in $N^{th}$ order Lovelock theory is pure Lovelock flat and gravity is kinematic in all critical $d=2N+1$ dimensions \cite{Ghosh}, giving the black hole solutions only in $d\geq 2N+2$ dimensions \cite{Dadhich}. These black hole solutions in $d=3N+1$ dimensions are indistinguishable from black holes in Einstein's gravity as for all of them the gravitational potential goes as $1/r$ \cite{Sumanta}. This motivates the study of pure Lovelock black holes.

In section \ref{The LUA}, we setup LUA trajectories in the background of a spherically symmetric Schwarzschild type black hole spacetime and present general solutions of a LUA trajectory in section \ref{Set up of LUA}. We summarize the acceleration bounds for LUA trajectories in the background of Schwarzschild black hole in section \ref{LUA in Sch}. Using the techniques applied in the case of Schwarzschild spacetime, we calculate the corresponding acceleration bounds for radial LUA trajectories in the background of the three black hole geometries in three different theories of gravity, namely the background geometries of charged black hole in Einstein's gravity in $d$ dimensions with $d\geq 4$ in section \ref{LUA in RN}, pure Lovelock black hole in section \ref{LUA in Pure LL} and black hole in $\mathcal{F}(R)$ theory in section \ref{LUA in f(R)}. Comparing the results for the unstable circular photon orbits  obtained in section \ref{The Shadow}  and the LUA trajectories in section \ref{The LUA}, we find an intriguing correspondence between the radius of the black hole shadow and the inverse of the acceleration bound of the LUA trajectories. Further the radius of photon sphere and the bound on the distance of closest approach for LUA trajectories are found to be equal. This is a fascinating correspondence since the notion of black hole shadow is completely independent of the LUA trajectories. To check the robustness of this equivalence, we further investigate this correspondence in a general Schwarzschild type spherically symmetric black hole spacetime in section \ref{The correspondence in Sch type BHs}. A more general class of spherically symmetric black hole spacetimes are explored in section \ref{The correspondence in AB type BHs} for such correspondence. The conclusions and discussions are presented in the last section.  

The signature adopted is $(+,-,-,-)$ and the natural units, $k_B=c=G=\hbar=1$ are used throughout the paper.

\section{Black hole shadow parameters}\label{The Shadow}

Two of the important notions related to the shadow of a static spherically symmetric black hole are the photon sphere and the critical impact parameter. The family of null trajectories with the value of impact parameter being infinitesimally larger than that of the critical impact parameter are the last set of photon trajectories which can just spiral outwards from an unstable circular orbit on the photon sphere and define the boundary of the shadow. For the Schwarzschild metric, these two notions are well studied \cite{Synge,Luminet} and have known values in terms of the mass $M$ of the black hole as, the radius of the photon sphere $r_{ph}=3 M$ and the critical impact parameter $b_{cr}=3\sqrt{3} M$. In the conventional approach, the radius of photon sphere is obtained from the peak of the effective potential encountered by photons moving in the background spacetime. For some static spherically symmetric black holes, there may exist several photon spheres, in which case, it is the unstable photon sphere which is relevant for the critical light rays spiralling outwards and it must be distinguished from the stable photon sphere about which the perturbed light trajectory may just oscillate. Another equivalent geometric approach uses the notions of geodesic curvature $k_g(r)$ and Gauss curvature $\mathcal{K}(r)$ in the context of optical metric of black holes \cite{Li,Qiao}. For spherically symmetric black holes, the unstable photon sphere satisfies the conditions $k_g(r)=0$ and $\mathcal{K}(r)<0$ while the stable photon sphere satisfies $k_g(r)=0$ and $\mathcal{K}(r)>0$.

In following sections, for the sake of completeness, we set up the basic equations leading to the radii of photon spheres and the shadow radius for static spherically symmetric black hole spacetimes of the Schwarzschild type and then solve them for particular solutions, namely the charged black hole in Einstein's gravity in $d$ dimensions, pure Lovelock black hole in $d$ dimensions and the black hole in $\mathcal{F}(R)$ gravity with conformally coupled scalar field in four dimensions. In case of the charged black hole in Einstein's gravity and the black hole in $\mathcal{F}(R)$ gravity, we encounter two photon spheres, one of which is unstable with respect to the radial perturbations and hence used to obtain the shadow radius.

\subsection{Basic equations and definitions}\label{Basics of shadow}

Consider a $d$-dimensional static spherically symmetric and asymptotically flat black hole metric,
\begin{eqnarray}
{ds}^{2} &=& f(r) \, {dt}^{2}-{f(r)}^{-1}{dr}^{2}-r^{2}\,d\Omega_{d-2}^{2}
\label{general metric}
\end{eqnarray}
where $f(r)$ is a smooth differentiable function with $f(r_H)=0$ at some radius $r_H$ which is the outer horizon of the black hole and $f(r) \rightarrow 1 $ as $r \rightarrow \infty $. Owing to the spherical symmetry, it is sufficient to consider the null geodesic motion in an equatorial plane. A photon moving in an equatorial plane will have only three non zero velocity components. The conserved quantities along the null geodesics corresponding to the two Killing vectors $\partial_t$ and $\partial_{\phi}$ lead to the two constants of motion,
\begin{equation}
f(r) \, \dot{t} = E \quad\text{and}\quad r^{2}\,\dot{\phi} = L
\end{equation}
where $E$ and $L$ are the energy and angular momentum of the photon as measured by an observer at infinity. Using these in the radial geodesic equation, the first integral for geodesic motion of photons can be written as
\begin{eqnarray}
\frac{\dot{r}^2}{L^2} + \frac{f(r)}{r^2} &=& \frac{1}{b^2}\label{null geodesic}
\end{eqnarray}
where $b=L/E$ is the impact parameter for the trajectory. The second term on left hand side is the effective potential where we define,
\begin{eqnarray}
V_{eff}(r) &=& \frac{f(r)}{r^2} \label{veff}
\end{eqnarray}
The effective potential plays an important role in calculating the size of the black hole shadow. The asymptotic flatness of metric at spatial infinity implies that $V_{eff}(r)$ falls as $1/r^2$ as $r \rightarrow \infty $ while at the outer horizon $r_H$ we have $V_{eff}(r_H) = 0 $ since $f(r_H)=0$. Thus $V_{eff}(r)$ has atleast one maxima between $r_H$ and $ r \rightarrow \infty$. This unstable extremum or the in the case of multiple unstable extrema, the one with the largest value of the potential is relevant for the shadow of the black hole as seen by an observer far away from the black hole. Here we assume, for the sake of simplicity, that all the sources of light are placed uniformly at infinity. In case, there are sources between the outer event horizon and the radius of unstable extremum, then the nature of the shadow as seen by the observer at infinity would be different and would depend on the distribution of these light sources.

The relevant unstable extremum point $r_{ph}$ described above leads to a circular orbit for photons and the value of the corresponding impact parameter, called as the critical impact parameter $b_{cr}$, then gives us the boundary of a black hole shadow. The light ray which just escapes the circular orbit at $r_{ph}$ and reaches the observer at distance $r_0$ can be projected straight backwards from $r_0$ to get the radius of the black hole shadow as
\begin{eqnarray}
    r_{sh} &=& r_0^2\frac{d\phi}{dr} \Bigg|_{r=r_0}  = \left( \frac{1}{b_{cr}^2} - V_{eff}(r_0)\right)^{-\frac{1}{2}}
\end{eqnarray}
which for the observer at infinity, $r_0 \rightarrow \infty$ leads to the critical impact parameter $r_{sh} = b_{cr}$. The critical impact parameter $b_{cr}$ can be obtained by setting $r = r_{ph}$ with $\dot{r}=0$ in Eq.(\ref{null geodesic}) to get 
\begin{eqnarray}
b_{cr} &=& \frac{1}{\sqrt{V_{eff}(r_{ph})}}
\end{eqnarray}
This corresponds to the turning point $r_{ph}$ for which the effective potential is maximum. For the effective potential in Eq. (\ref{veff}), the photon sphere radius can be obtained as a solution to the equation,
\begin{eqnarray}
0 &=&\frac{d V_{eff}}{dr}\Bigg|_{r=r_{ph}} =\frac{r_{ph}\, f'(r_{ph})-2\, f(r_{ph})}{r_{ph}^3}.\label{photon sphere}
\end{eqnarray}
where, the prime denotes the derivative with respect to radial coordinate $r$. Using the radius of photon sphere $r_{ph}$, it is then straightforward to get the critical impact parameter $b_{cr}$ which gives the shadow radius $r_{sh}$ of black hole as,
\begin{eqnarray}
r_{sh} &=& b_{cr} = \frac{1}{\sqrt{V_{eff}(r_{ph})}}= \frac{2\sqrt{f(r_{ph})}}{f'(r_{ph})}\label{shadow radius}
\end{eqnarray}
where the last equality is obtained using the equation for the photon sphere radius Eq.(\ref{photon sphere}). Thus, given a specific form of $f(r)$, the size of the shadow of any black hole of the form of Eq.(\ref{general metric}) can be evaluated using Eq.(\ref{photon sphere}) and Eq.(\ref{shadow radius}). We use these equations in the next sections to solve for the shadow radius of the black hole under consideration.

\subsection{Shadow of a $d$-dimensional charged black hole in Einstein's gravity}\label{Shadow of RN}

The metric of $d$-dimensional charged black hole in Einstein's gravity has the form as given in Eq.(\ref{general metric}) with the metric component $f(r)$ given by \cite{Tang},
\begin{eqnarray}
f(r) &=& 1-\frac{2M}{r^{d-3}}+\frac{Q^2}{r^{2(d-3)}}\label{metric for RN BH}
\end{eqnarray}
where $M$ is the mass and $Q$ is the charge of the black hole with $M>Q$ and $d\geq 4$. The black hole has two horizons. The outer horizon of the black hole is at radius,
\begin{eqnarray}
r_H &=& \left(M+\sqrt{M^2-Q^2}\right)^{\frac{1}{d-3}}
\end{eqnarray}
Substituting the metric function $f(r)$ in Eq.(\ref{photon sphere}) and simplifying we get the equation defining the photon sphere radius as,
\begin{eqnarray}
r_{ph}^{2(d-3)}-M(d-1)\,r_{ph}^{d-3}+(d-2) \,Q^2 &=& 0
\end{eqnarray}
According to the rule of signs, this equation will have two positive real roots, i.e, the black hole will have two photon spheres. The radii of these photon spheres are given by,
\begin{eqnarray}
r_{ph} &=& \left(\frac{M(d-1)\pm\sqrt{M^2 (d-1)^2-4(d-2)Q^2}}{2}\right)^{\frac{1}{d-3}}\label{photon sphere of RN BH}
\end{eqnarray}
where photon sphere corresponding to `$+$' sign is an unstable photon sphere while the one corresponding to `$-$' sign is stable with respect to radial perturbations. The light rays forming the boundary of the shadow spiral towards the unstable photon sphere. Hence to obtain the shadow of the black hole we use solution of the outer unstable photon sphere in Eq.(\ref{shadow radius}). This gives the radius of shadow to be equal to,
\begin{eqnarray}
r_{sh} &=& r_{ph} \sqrt{\frac{(d-2)\,r_{ph}^{d-3}}{(d-3)\,\left(r_{ph}^{d-3}-M\right)}}
\label{shadow of RN BH}
\end{eqnarray}

Comparing the expressions for the radius of horizon $r_{H}$, the radius of unstable photon sphere $r_{ph}$ and the shadow radius $r_{sh}$ and using the constraint $M>Q$, one can show that the inequalities $r_H<r_{ph}<r_{sh}$ always hold for all $d\geq 4$. Also, by specifying some values of $M$ and $Q$ with $M>Q$, one can check that all the radii $r_{H}$, $r_{ph}$ and $r_{sh}$ decrease with increasing dimension $d$, i.e, a black hole with same mass $M$ and charge $Q$ in higher dimension will have smaller horizon, photon sphere and shadow compared to the ones in lower dimensions.

For consistency, one can check that in four dimensions, the results for the photon sphere radius and the shadow radius simplify to,
\begin{eqnarray}
r_{ph} =\frac{3M+\sqrt{9M^2-8Q^2}}{2} &\text{and}& r_{sh}=\sqrt{\frac{\left(3M+\sqrt{9M^2-8Q^2}\right)^3}{2\left(M+\sqrt{9M^2-8Q^2}\right)}}
\end{eqnarray}
which are the corresponding values of the Reissner–Nordström black hole \cite{Zakharov 1,Zakharov 2}.

\subsection{Shadow of a pure Lovelock black hole}\label{Shadow of Pure LL}

The pure Lovelock theory of $N^{th}$ order admits an interesting class of pure Lovelock charged black hole solutions. These solutions posses a characterizing property of pure Lovelock theory of having the universal character of their thermodynamic parameters in terms of the event horizon radius \cite{Dadhich}. The metric of a $d$-dimensional $N^{th}$ order pure Lovelock charged black hole takes the form of Eq.(\ref{general metric}) with the function $f(r)$ given by \cite{Naresh},
\begin{eqnarray}
f(r) &=& 1-\left(\frac{2^N M}{r^{d-2N-1}}-\frac{Q^2}{r^{2d-2N-4}}\right)^{\frac{1}{N}}
\end{eqnarray}
where $M$ is the mass and $Q$ is the charge of the black hole with $d\geq 2N+2$. Substituting the metric function $f(r)$ in the equation of photon sphere, Eq.(\ref{photon sphere}) we get,
\begin{eqnarray}
r_{ph}^{2d-2N-4} \left[2^N M\, r_{ph}^{d-3}-Q^2\right]^{N-1}-\left[\frac{2^N M(d-1)\, r_{ph}^{d-3}-2(d-2)Q^2}{2N}\right]^N &=& 0 \quad \quad
\end{eqnarray}
This equation is not analytically solvable in general. Hence, we solve for a special case, an uncharged black hole with $Q=0$, which is the vacuum solution of the pure Lovelock theory. The metric function in this case becomes,
\begin{eqnarray}
f(r) &=& 1-\left(\frac{2^N M}{r^{d-2N-1}}\right)^{\frac{1}{N}}\label{metric for pure LL BH}
\end{eqnarray}
This black hole has a horizon at radius,
\begin{eqnarray}
r_H &=& 2^{^{\frac{N}{(d-2N-1)}}}\,\, M^{^{\frac{1}{(d-2N-1)}}}
\end{eqnarray} 
Further with $Q=0$, the equation for photon sphere radius simplifies to,
\begin{eqnarray}
r_{ph}^{d-2N-1}-M \left(\frac{d-1}{N}\right)^N &=& 0 
\end{eqnarray}
This equation will have only one positive real root by the Descartes rule of sign for the above polynomial. Thus, in the case of uncharged pure Lovelock black hole we encounter only one photon sphere and it's radius is given by,
\begin{eqnarray}
r_{ph} &=& \left(\frac{d-1}{N}\right)^{\frac{N}{d-2N-1}} M^{^{\frac{1}{d-2N-1}}}\label{photon sphere of pure LL BH}
\end{eqnarray}
Using this expression for photon sphere radius in  Eq.(\ref{shadow radius}), we obtain the shadow radius of the uncharged Lovelock black hole to be,
\begin{eqnarray}
r_{sh} &=& \sqrt{\frac{d-1}{d-2N-1}}\,\, r_{ph} \label{shadow of pure LL BH}
\end{eqnarray}
Comparing the radii of horizon, photon sphere and shadow, it can be easily shown that the inequalities
\[r_H<r_{ph}<r_{sh}\]
always hold for all $d\geq 2N+2$. In Einstein's gravity i.e. for $N=1$, one can check that the expressions reduce to,
\begin{equation}
r_{ph} = (d-1)^{\frac{1}{d-3}}\,M ^{\frac{1}{d-3}} \quad\quad\text{and} \quad\quad r_{sh}= \sqrt{\frac{d-1}{d-3}}\,(d-1)^{\frac{1}{d-3}}\,M ^{\frac{1}{d-3}}\label{photon sphere and shadow for Schwazrschild}
\end{equation} 
which are the radii of photon sphere and shadow of the $d$-dimensional Schwarzschild black hole.

For a given order $N$ of the Lovelock theory in the lowest possible dimension $d=2N+2$ for a black hole, the radius of photon sphere $r_{ph}$ and the shadow radius $r_{sh}$ given by Eq.(\ref{photon sphere of pure LL BH}) and (\ref{shadow of pure LL BH}), will always be proportional to the mass $M$ of the black hole. In such a case, the expressions in Eq.(\ref{photon sphere of pure LL BH}) and (\ref{shadow of pure LL BH}) simplify to,
\begin{eqnarray}
r_{ph} = \left(2N+1\right)^N \, M &\text{and} & r_{sh} = \left(2N+1\right)^{N+\frac{1}{2}} \, M
\end{eqnarray}
From these expressions one can easily see that the sizes of the photon sphere and the shadow of the black hole increase with the increasing order $N$ of the Lovelock theory. Thus a black hole in $4$-dimensional Einstein's gravity will have a smaller shadow than the one in $6$-dimensional pure Gauss-Bonnet gravity.

\subsection{Shadow of $\mathcal{F}(R)$ black hole}\label{Shadow of f(R)}

We consider a black hole solution of a $\mathcal{F}(R)$ gravity theory non-minimally coupled to a self-interacting scalar field $\Phi(r)$ \cite{Karakasis}. For this black hole solution the $\mathcal{F}(R)$ gravity was modelled with a non-linear curvature correction to the Ricci scalar $R$ as,
\begin{eqnarray}
\mathcal{F}(R) &=& R-2\alpha \sqrt{R}\label{model of f(R)}
\end{eqnarray}
where $\alpha$ is a model parameter with the dimensions of inverse length. This model with a conformally coupled scalar field has a asymptotically flat black hole solution in four dimensions given by the metric in Eq.(\ref{general metric}) with $d=4$ and the metric function \cite{Karakasis},
\begin{eqnarray}
f(r) &=& \frac{1}{2}-\frac{1}{3\alpha r}+\frac{3}{64\alpha^2 r^2}\label{metric for f(R)}
\end{eqnarray}
with $\alpha >0$. The model parameter $\alpha$ is the only free parameter for the black hole and it is inversely proportional to the mass of the black hole. The black hole has two horizons given by,
\begin{eqnarray}
r_{\pm} &=& \frac{8\pm \sqrt{10}}{24\alpha}
\end{eqnarray}
The corresponding scalar field solution is \cite{Karakasis},
\begin{eqnarray}
\Phi(r) &=& \frac{\sqrt{6}}{1-4\alpha r}
\end{eqnarray}
which diverges at a radius $r_d=1/4\alpha$ that lies between the inner horizon $r_{-}$ and the outer horizon $r_+$. In region of our interest, outside the outer horizon $r_+$, both the metric function $f(r)$ and the scalar field $\Phi(r)$ are finite. The unstable photon sphere responsible for shadow lies in this region. The equation of a photon sphere, Eq.(\ref{photon sphere}) for this black hole becomes,
\begin{eqnarray}
16 \,\alpha^2\, r_{ph}^2-16\,\alpha\, r_{ph}+3 &=& 0
\end{eqnarray}
According to the rule of signs, this equation will have two positive real roots, that is the black hole has two photon spheres. The unstable photon sphere outside the horizon $r_+$ has radius equal to,
\begin{eqnarray}
r_{ph} &=& \frac{3}{4\alpha}\label{photon sphere of f(R)}
\end{eqnarray}
Substituting this radius of the unstable photon sphere in Eq.(\ref{shadow radius}) we get the radius of the black hole shadow as,
\begin{eqnarray}
r_{sh} &=& \frac{6}{\sqrt{5}} r_{ph}\label{shadow of f(R)}
\end{eqnarray}
For $\alpha>0$, the horizon, the photon sphere and the shadow radii will always satisfy the relation $r_+<r_{ph}<r_{sh}$. From the expressions, one can easily show that with the increasing value of $\alpha$, i.e, the increasing contribution from non-linear curvature correction term to the gravity, the sizes of horizon, photon sphere and shadow decrease.

\section{Acceleration bounds for LUA trajectories}\label{The LUA}

A linearly uniformly accelerated (LUA) trajectory is the curved spacetime generalization of the hyperbolic Rindler trajectory in flat spacetime. The Rindler trajectories in flat spacetime can be quantified in a covariant definition in terms of the Letaw-Frenet equations wherein, out of the three geometric scalars for a curve, only the curvature is a non-zero constant and equal to the magnitude of acceleration whereas the torsion and hyper-torsion scalars vanish \cite{Letaw}. Generalising this covariant construction to curved spacetime ensures that the corresponding curves with constant curvature scalar and vanishing torsion and hyper-torsion are locally hyperbolic and linear in a local inertial frame at any event along the curve \cite{Kolekar}.

In the following sub-sections, we first recall the general setup to obtain the LUA trajectories in static spherically symmetric black hole spacetimes of Schwarzschild type and then briefly summarise the results of radial LUA trajectories in Schwarzschild geometry regarding the bounds on acceleration and the distance of closest approach. We then follow the same procedure to calculate the corresponding bounds for radial LUA trajectories in the background of a charged black hole in Einstein's gravity in $d$ dimensions, a  pure Lovelock black hole and the $\mathcal{F}(R)$ black hole considered in section \ref{The Shadow}.

\subsection{General setup}\label{Set up of LUA}

The linearity condition imposed by setting torsion and hypertorsion to zero in the Letaw-Frenet equations for uniform acceleration leads to the following expression \cite{Kolekar},
 \begin{eqnarray}
w^{i}-{|a|}^{2}u^{i}=0
\label{linearity}
\end{eqnarray}
where  $w^{i}=u^{j}{\nabla}_{j}a^{i}$ and $a^i$ and $u^i$ are the acceleration and velocity vectors along the trajectory. All solutions to the above constraint equation along with constant magnitude of acceleration $|a|$ form the set of LUA trajectories for a given spacetime.

We consider a static spherically symmetric background metric of the Schwarzschild type of the form,
\begin{eqnarray}
{ds}^{2} &=& f(r) \, {dt}^{2}-{f(r)}^{-1}{dr}^{2}-r^{2}\,{d\theta}^{2}-r^{2}{\sin}^{2}\theta \, {d\phi}^{2}.
\end{eqnarray}
where the function $f(r)$ is a metric function defining a black hole spacetime with a horizon at some radius $r_H$ such that $f(r_H)=0$ which we consider to be the outer horizon of the black hole and $f(r) \rightarrow 1 $ as $r \rightarrow \infty $. We next consider a radial trajectory with all angular co-ordinates fixed and solve for the LUA trajectory consistent with the constraint Eq.(\ref{linearity}). We obtain the solution in terms of the components of velocity vector of the trajectory \cite{Kajol} as,
\begin{eqnarray}
\frac{dt}{d\tau}&=&{f(r)}^{-1}\left(|a|r+h\right)\label{time velocity}\\
\frac{dr}{d\tau}&=&\pm\sqrt{\left(|a|r+h\right)^2-f(r)}\label{radial velocity}\\
\frac{d\theta}{d\tau} &=& \frac{d\phi}{d\tau} =0
\end{eqnarray}
where, $\tau$ is the proper time along the trajectory, $|a|$ is the uniform magnitude of acceleration and $h$ is specified as boundary data. In flat spacetime, the constant $h$ parametrizes the shifted Rindler wedges and form a symmetry of Minkowski spacetime corresponding to the global translation Killing vector $\partial_X$. Although the translation symmetry does not hold for spherically symmetric spacetimes with a black hole at centre, the role of $h$ for generating a shift in the radial LUA trajectory with respect to boundary events on the future and past null infinity still holds. A more detailed discussion is presented in \cite{Kajol,Sanved}. Given an explicit form of the metric function $f(r)$ and suitable boundary condition, one can obtain a particular solution for the radial LUA trajectory.

These results can be extended to a $d$-dimensional Schwarzschild type spacetime of the form of Eq.(\ref{general metric}). In such a case, the components $\Gamma^k_{ij}$ of Christoffel connection vanish for all $k\neq ({r,t})$ and $(i,j)=(r,t)$. Then the only non zero components of acceleration $a^i$ and the vector $w^i$ are along the radial and temporal directions. This leads to the exact same linearity condition as in the four dimensional spacetime for radial LUA trajectories. Hence the solutions in Eq.(\ref{time velocity}) and (\ref{radial velocity}) are valid also for radial LUA trajectories in $d$-dimensional Schwarzschild type black hole spacetimes defined by Eq.(\ref{general metric}).

\subsection{LUA trajectory in the Schwarzschild spacetime}\label{LUA in Sch}

We briefly summarise the results in \cite{Kajol,Sanved} pertaining to the radial LUA trajectory and the acceleration bounds in Schwarzschild spacetime. For the Schwarzschild metric,
\begin{eqnarray}
{ds}^{2} &=& \left(1-\frac{2M}{r}\right) \, {dt}^{2}-{\left(1-\frac{2M}{r}\right)}^{-1}{dr}^{2}-r^{2}\,{d\theta}^{2}-r^{2}{\sin}^{2}\theta \, {d\phi}^{2}
\label{Schwarzschild metric}
\end{eqnarray}
where $M$ is the mass of the black hole, the solution for the radial LUA trajectory in Eqs.(\ref{time velocity}) and (\ref{radial velocity}) become,
\begin{eqnarray}
\dfrac{dt}{d\tau}&=&{\left(1-\frac{2M}{r}\right)}^{-1}\left(|a|r+h\right)\label{Schwarzschild time velocity}\\
\dfrac{dr}{d\tau}&=&\pm|a| \, \sqrt{\frac{(r-r_{min})(r-r_{max})(r-r_n)}{r}}\label{Schwarzschild radial velocity}
\end{eqnarray}
where $r_{min}$, $r_{max}$ and $r_n$ are the roots of the cubic polynomial, $r\,{\left(|a| r+h\right)}^2 -r+2M$. The roots $r_{min}$ and $r_{max}$ are the turning points of the incoming and outgoing trajectories starting from radius $r>r_{min}$ and $r<r_{max}$ respectively. The root $r_n$ being negative does not have any physical significance. For radial LUA trajectories initially moving towards the black hole starting from the radial infinity, the turning points $r_{min}$ and $r_{max}$ were found to be positive real only upto a certain maximum value of the magnitude of acceleration $|a|$ written as a bound value $|a|_b$. For $|a| > {|a|}_{b}$ the turning points become complex and the ingoing LUA trajectory always falls into the horizon. The bound on acceleration can be expressed as a function of the mass of the black hole and the initial boundary data $h$ as,
\begin{eqnarray}
{|a|}_{b} &=& \frac{\left(-9 \, h + h^3 + \sqrt{{(3+h^2)}^3} \right)}{27 \,M }\label{Schwarzschild acceleration bound}
\end{eqnarray}
The turning point $r_{min}$ corresponding to the bound value of acceleration gives the lower bound on the distance of closest approach $r_{b}$ as,
\begin{eqnarray}
{r}_{b} &=& \frac{2}{3 {|a|}_b}\left(\frac{\sqrt{3+h^2}}{2}-h\right)\label{Schwarzschild closest approach}
\end{eqnarray}
At the bound $|a|=|a|_b$ the two turning points $r_{min}$ and $r_{max}$ coincide and give the lower bound on the turning point $r_{min}$ equal to $r_b$ for the trajectory to escape to infinity. Thus at the bound the cubic polynomial $r\,{\left(|a| r+h\right)}^2 -r+2M$ has a double root. This gives an algebraic condition on the bound that, for the magnitude of acceleration equal to the bound value $|a|_b$, both the cubic polynomial and its derivative with respect to the radial coordinate $r$ must vanish simultaneously at $r=r_b$. We use this equivalent condition henceforth in the following sections to find the bounds on the magnitude of acceleration $|a|_b$ and the distance of closest approach $r_b$ for the LUA trajectories in the specific black hole spacetimes, namely the charged black hole in Einstein's gravity in $d$ dimensions, the pure Lovelock black hole and the $\mathcal{F}(R)$ black hole spacetime. Another perspective to understand the existence of an acceleration bound for a radial LUA trajectory in the Schwarzschild spacetime is to compare it with the acceleration of an stationary observer's trajectory at fixed $r$ \cite{Sanved}. The bound value $|a|_b$ is equal to the magnitude of acceleration of a stationary trajectory at the turning point $r=r_b$, beyond which the inward gravity of the black hole dominates the outward acceleration of the LUA trajectory, and so the trajectory falls into the horizon. 

The special case of initial boundary data $h=0$ in flat spacetime corresponds to family of LUA trajectories vis-a-vis the hyperbolic Rindler trajectories in flat spacetime with $|a| \in (0,\infty]$ belonging to a Rindler wedge with the bifurcation point at the origin $(T,X)=(0,0)$ of the Rindler past and future horizons and which asymptote to $T-X=0 $ and $T+X=0$ at future and past null infinity. In the Schwarzschild spacetime, the $h=0$ case does fix a set of family of LUA trajectories however they do not belong to a single Rindler wedge since for all different values of $|a|$ the asymptotes of the LUA trajectories at future and past null infinity are different. In this case, the allowed values of acceleration are $|a| \in (0,|a|_b ]$ with the upper bound on the magnitude of acceleration, Eq.(\ref{Schwarzschild acceleration bound}) and the distance of closest approach, Eq.(\ref{Schwarzschild closest approach}) simplified to,
\begin{equation}
|a|_b = \frac{1}{3\sqrt{3} M} \quad\quad \text{and} \quad\quad r_b= 3M =\frac{1}{\sqrt{3}\, |a|_b}
\end{equation}
We shall later show that it is this set of LUA trajectories with $h=0$ which have an interesting connection with the shadow parameters in the same black hole spacetime geometries.

The acceleration bounds on LUA trajectories are consequences of the presence of a black hole horizon. In flat spacetime, the past and future Rindler horizons and hence the Rindler quadrant are independent of acceleration $|a|$ of the trajectory. All the Rindler trajectories with acceleration $|a|$ in the range $0<|a|<\infty$ lie in the same Rindler quadrant. However, in the case of black hole spacetime, the turning point of a LUA trajectory, the corresponding Rindler horizons and their bifurcation point are all functions of $|a|$ and are different for trajectories with different acceleration $|a|$. Increasing the value of $|a|$ of LUA trajectory not only decreases the value of the turning point, as in the case of flat spacetime, but also the value of the bifurcation point. The lowest possible value of the bifurcation point is limited by the horizon of the black hole which in turn limits the turning point to the bound $r_b$ and the acceleration to $|a|_b$. This is explained in detail in our earlier work \cite{Sanved} in section III C. In non black hole spacetimes, there is no horizon to limit the bifurcation point of the past and future acceleration horizons and hence the acceleration bounds do not exist.

\subsection{Bounds in $d$-dimensional charged black hole in Einstein's gravity}\label{LUA in RN}

For a $d$-dimensional charged black hole in Einstein's gravity, the metric function $f(r)$ is given by Eq.(\ref{metric for RN BH}). Substituting this metric function in Eq.(\ref{radial velocity}) with $h=0$, we get the solution for the radial LUA trajectories to be,
\begin{eqnarray}
\left(\frac{dr}{d\tau}\right)^2 &=& {|a|}^2r^2- 1+\frac{2M}{r^{d-3}}-\frac{Q^2}{r^{2(d-3)}}
\end{eqnarray}
Beyond the bound value of acceleration, the turning point of the trajectory becomes complex valued. So to get the bound on acceleration, we first write the equation of the turning point of the trajectory. At the turning point, radial velocity is momentarily zero. Hence the turning point $r_t$ of the LUA trajectory will satisfy,
\begin{eqnarray}
|a|^2 r_t^{2(d-2)}-r_t^{2(d-3)}+ 2M r_t^{d-3} -Q^2 &=& 0\label{turning point in RN}
\end{eqnarray}
Applying the Descartes rule of sign to this polynomial, we can say that it will have either three positive real roots or one positive real root. For acceleration less than the bound value we expect three positive real roots, out of which two will coincide exactly at the bound value. These two roots will become complex valued for $|a| > |a|_b$ and then the polynomial will have only one real positive root. At the bound $|a|=|a|_b$, the turning point $r_t=r_b$ is an extremum of the above polynomial and hence it also satisfies,
\begin{eqnarray}
0 &=& \left[\frac{d}{dr}\left(\,\,|a|^2 r^{2(d-2)}-r^{2(d-3)}+ 2M r^{d-3} -Q^2\,\,\right)\right]_{|a|=|a|_b,\, r=r_b}\nonumber\\
&=& |a|_b^2 (d-2)\, r_b^{2d-5} -(d-3)\, r_b^{2d-7} +(d-3)M\, r_b^{d-4}\label{merging of roots in RN}
\end{eqnarray}
Solving the two equations, Eq.(\ref{turning point in RN}) and Eq.(\ref{merging of roots in RN}) simultaneously for $r_b$ and $|a|_b$ we get,
\begin{eqnarray}
r_b &=& \left(\frac{M(d-1)+\sqrt{M^2 (d-1)^2-4(d-2)Q^2}}{2}\right)^{\frac{1}{d-3}}\label{distance of closest approach in RN}\\
\text{and} \quad |a|_b &=& \frac{1}{{r_b}^{d-2}} \sqrt{(d-3)(M {r_{ph}}^{d-3}-Q^2)}\nonumber\\
&=& \frac{1}{r_{b}}\sqrt{\frac{(d-3)\,({r_{b}}^{d-3}-M)}{(d-2)\,{r_{b}}^{d-3}}} \label{acceleration bound in RN}
\end{eqnarray}
where the last equality is obtained using the equation for turning point, Eq.(\ref{turning point in RN}). We note here that the bound on acceleration $|a|_b$ and $r_b$ are not simply inversely proportional to each other as in the case of Schwarzschild black hole. The additional charge on the black hole leads to a more complex relation between the two bounds. One can note that expression for the acceleration bound $|a|_b$ in Eq.(\ref{acceleration bound in RN}) and bound on distance of closest approach $r_b$ in Eq.(\ref{distance of closest approach in RN}) are identical to the expressions for the shadow radius $r_{sh}$ and the photon sphere radius $r_{ph}$ given in Eqs.(\ref{shadow of RN BH}) and (\ref{photon sphere of RN BH}) in section \ref{Shadow of RN}.

\subsection{Bounds in a pure Lovelock black hole spacetime}\label{LUA in Pure LL}

We consider an uncharged pure Lovelock black hole defined by the metric in Eq.(\ref{general metric}) with the function $f(r)$ given by Eq.(\ref{metric for pure LL BH}).
The solution to the radial LUA trajectory with initial data $h=0$ in the background of this black hole using Eq.(\ref{radial velocity}) becomes,
\begin{eqnarray}
\left(\frac{dr}{d\tau}\right)^2 &=& {|a|}^2r^2- 1+\left(\frac{2^N M}{r^{d-2N-1}}\right)^{\frac{1}{N}}
\end{eqnarray}
The turning point $r_t$ of this trajectory can be obtained by setting the left hand side of the above equation to zero to get,
\begin{eqnarray}
{|a|}^2 r_t^{\frac{d-1}{N}}- r_t^{\frac{d-2N-1}{N}}+2 M^{^{\frac{1}{N}}} &=& 0 \label{turning point in pure LL}
\end{eqnarray}
According to the Descartes rule of sign, this polynomial will have two real positive roots or no real positive root. The two real positive roots are similar to the ones in the case of Schwarzschild black hole. These two roots coincide at the bound value of the magnitude of acceleration, beyond which no real turning point exists and the trajectory always falls into the horizon. Hence for the trajectory with acceleration equal to the bound value, the derivative of the above polynomial with respect to radial coordinate $r$ will be zero at the turning point $r=r_b$. Thus in addition to Eq.(\ref{turning point in pure LL}), the bound values $|a|_b$ and $r_b$ satisfy the condition,
\begin{eqnarray}
0&=&\left[\frac{d}{dr}\left({|a|}^2 r^{\frac{d-1}{N}}- r^{\frac{d-2N-1}{N}}+2 M^{^{\frac{1}{N}}}\right)\right]_{|a|=|a|_b,\, r=r_b} \nonumber\\
&=& {|a|_b}^2\left(\frac{d-1}{N}\right) r_b^{\frac{d-1}{N}-1}- \left(\frac{d-2N-1}{N}\right) r_b^{\frac{d-1}{N}-3}\label{merging of roots in pure LL}
\end{eqnarray}
Solving Eq.(\ref{turning point in pure LL}) and (\ref{merging of roots in pure LL}) simultaneously for the bounds $|a|_b$ and $r_b$ we get,
\begin{eqnarray}
r_b &=& \left(\frac{d-1}{N}\right)^{\frac{N}{d-2N-1}} M^{^{\frac{1}{d-2N-1}}}\label{distance of closest approach in pure LL}\\
\text{and} \quad{|a|}_b &=& \sqrt{\frac{d-2N-1}{d-1}}\,\,\frac{1}{r_b}\label{acceleration bound in pure LL}
\end{eqnarray}
Here, the bound on the acceleration $|a|_b$ is inversely proportional to the bound on the distance of closest approach $r_b$ as in the in case of the Schwarzschild black hole with the proportionality constant varying with the dimension $d$ of the spacetime and the order $N$ of the theory. One can note that expression for the acceleration bound $|a|_b$ in Eq.(\ref{acceleration bound in pure LL}) and bound on distance of closest approach $r_b$ in Eq.(\ref{distance of closest approach in pure LL}) are identical to the expressions for the shadow radius $r_{sh}$ and the photon sphere radius $r_{ph}$ given in Eqs.(\ref{shadow of pure LL BH}) and (\ref{photon sphere of pure LL BH}) in section \ref{Shadow of Pure LL}.

\subsection{Bounds in $\mathcal{F}(R)$ black hole spacetime}\label{LUA in f(R)}

We consider the LUA trajectories in the four dimensional black hole spacetime defined by the metric function $f(r)$ given in Eq.(\ref{metric for f(R)}) in the $\mathcal{F}(R)$ gravity conformally coupled to a scalar field. The solution for a LUA trajectory in this background spacetime with the initial data $h=0$ becomes,
\begin{eqnarray}
\left(\frac{dr}{d\tau}\right)^2 &=& {|a|}^2r^2-\frac{1}{2}+\frac{1}{3\alpha r}-\frac{3}{64\alpha^2 r^2}
\end{eqnarray}
At the turning point $r_t$ the radial velocity vanishes and the turning point will satisfy, 
\begin{eqnarray}
192\, \alpha^2\, |a|^2\, r_t^4 - 96\, \alpha^2\, r_t^2 + 64\,\alpha\, r_t -9 &=& 0\label{turning point in f(R)}
\end{eqnarray}
According to the rule of signs, this equation will have three real positive roots or one real positive root, as in the case of the charged black hole in Einstein's gravity. At the acceleration equal to the bound value, two of the three roots merge giving the bound on the distance of closest approach $r_b$, which thus satisfies the equation,
\begin{eqnarray}
0 &=& \frac{d}{dr} \left( 192\, \alpha^2\, |a|^2\, r^4-96\, \alpha^2\, r^2 +64\,\alpha\, r-9\right)\bigg|_{|a|=|a|_b,\, r=r_b}\nonumber\\
&=& 12\,\alpha\, |a|_b^2\, r_b^3-3 \,\alpha\, r_b+1 \label{merging of roots in f(R)}
\end{eqnarray}
Solving the above two equations, Eq.(\ref{turning point in f(R)}) and Eq.(\ref{merging of roots in f(R)}) for the bounds $|a|_b$ and $r_b$, we get,
\begin{eqnarray}
r_b &=& \frac{3}{4\alpha} \label{distance of closest approach in f(R)}\\
\text{and} \quad |a|_b &=& \frac{\sqrt{5}}{6} \frac{1}{r_b}\label{acceleration bound in f(R)}
\end{eqnarray}
In this case too the acceleration bound $|a|_b$ and the bound on the distance of closest approach $r_b$ are inversely proportional to each other. One can note that expression for the acceleration bound $|a|_b$ in Eq.(\ref{acceleration bound in f(R)}) and bound on distance of closest approach $r_b$ in Eq.(\ref{distance of closest approach in f(R)}) are identical to the expressions for the shadow radius $r_{sh}$ and the photon sphere radius $r_{ph}$ given in Eqs.(\ref{shadow of f(R)}) and (\ref{photon sphere of f(R)}) in section \ref{Shadow of f(R)}.

\section{Relation between shadow parameters and acceleration bounds for Schwarzschild type spacetimes}\label{The correspondence in Sch type BHs}

From the expressions obtained for the shadow parameters in section \ref{The Shadow} and for the acceleration bounds of the LUA trajectories in section \ref{The LUA}, one can see an interesting connection. The radii of photon sphere $r_{ph}$ and shadow radii $r_{sh}$ as seen by an observer far away from the black holes are directly related to the acceleration bounds $|a|_b$ and bounds on the closest approach $r_b$ for LUA trajectories,  in all four black hole spacetime geometries investigated, namely, the Schwarzschild black hole, the $d$-dimensional charged black hole in Einstein's gravity, the uncharged pure Lovelock black hole and the four dimensional black hole in the $\mathcal{F}(R)$ gravity. In particular we get the following relations,
\begin{equation}
r_{sh}=\frac{1}{|a|_b} \quad\quad \text{and} \quad\quad r_{ph}= r_b \label{correspondence}
\end{equation}
The relation is indeed intriguing since kinematics of the photon sphere and the shadow of a black hole involve null geodesics while the kinematics of the acceleration bounds involve timelike accelerated trajectories and both seem independent of each other. However, both the former and the latter, involve limiting cases in their own domain. Below, we explore these limiting cases in a general way. 

Consider a class of $d$-dimensional static spherically symmetric black hole spacetimes of Schwarzschild type defined by the form of the metric as in section \ref{The Shadow},
\begin{eqnarray}
{ds}^{2} &=& f(r) \, {dt}^{2}-{f(r)}^{-1}{dr}^{2}-r^{2}\,d\Omega_{d-2}^{2} \label{general class of spacetime}
\end{eqnarray}
where $f(r)$ is a smooth differentiable function with $f(r_H)=0$ at the horizon $r_H$ of the black hole and $f(r) \rightarrow 1 $ as $r \rightarrow \infty $. For such a black hole, the photon sphere radius and the shadow radius are given by Eq.(\ref{photon sphere}) and (\ref{shadow radius}) respectively. To relate these to the acceleration bounds, we first write the solution of a radial LUA trajectory with $h=0$ as given in Eq.(\ref{radial velocity}),
\begin{eqnarray}
\left(\frac{d r}{d\tau}\right)^2  &=&  |a|^2r^2 - f(r) \equiv  - \tilde{V}_{\mathit{eff}}(r)
\end{eqnarray}
with $|a| \neq 0$. The right hand side of this equation is a function of $r$ and can be interpreted as the negative of an effective potential $\tilde{V}_{\mathit{eff}}(r)$ for the motion, which determines the kinematics of the radial LUA trajectory. At the turning point of such a trajectory, the radial velocity is zero which constrains the effective potential to be zero at the turning point. A turning point, say $r_{t}$, thus satisfies the equation,
\begin{eqnarray}
0 &=& -\tilde{V}_{\mathit{eff}}(r_t) =|a|^2r_{t}^2-f(r_{t})  \label{turning point}
\end{eqnarray}
At the boundaries of our region of interest, i.e, at horizon and at spatial infinity, the first term $|a|^2r^2$ varies from $|a|^2r_H^2$ to $\infty$ whereas the second term $f(r)$ varies from $0$ to $1$. Therefore, in a background spacetime where a return LUA trajectory exists, these two terms are bound to coincide at at least two points between the horizon $r_H$ and radial infinity giving at least two positive real turning point solutions. The acceleration bound is then defined by the value of acceleration for which the two solutions coincide. Two equal roots of a polynomial are bound to be the roots of the polynomial and its derivative simultaneously. This gives two conditions that needs to be simultaneously satisfied by the bound on acceleration $|a|_b$ and the bound on the turning point $r_b$. The conditions are,
\begin{eqnarray}
\tilde{V}_{\mathit{eff}}(r)\bigg|_{|a|=|a|_b,\, r=r_b} &=& 0\label{constrain 1 on effective potential}\\
\text{and} \quad \frac{d\tilde{V}_{\mathit{eff}}(r)}{dr}\bigg|_{|a|=|a|_b,\, r=r_b} &=& 0 \label{constrain 2 on effective potential}
\end{eqnarray}
Writing these conditions in terms of the metric function $f(r)$, we get,
\begin{eqnarray}
|a|_b^2 r_{b}^2 - f(r_{b}) &=& 0 \label{potential=0}\\
2 |a|_b^2 r_b - f'(r_b) &=& 0 \label{derivative of potential=0}
\end{eqnarray}
Solving these two equations simultaneously, we get an expression for the acceleration bound $|a|_b$ in terms of the metric function $f(r)$ and a constraint equation for the bound on the turning point $r_b$ as,
\begin{eqnarray}
|a|_b^2 &=& \frac{f(r_b)}{r_b^2} \label{acceleration bound for general metric}\\
0 &=& 2 f(r_b) -r_b f'(r_b) \label{constraint equation for the bound on turning point}
\end{eqnarray}
The above constraint equation is same as that for the radius of photon sphere in Eq.(\ref{photon sphere}). Hence, the bound on distance of closest approach $r_b$ will always be equal to the radius of photon sphere $r_{ph}$ for all static spherically symmetric black hole spacetimes of the form of Eq.(\ref{general class of spacetime}). This further relates the bound on acceleration $|a|_b$ to the effective potential of the photon trajectories as,
\begin{eqnarray}
|a|_b &=& \sqrt{\frac{f(r_b)}{r_b^2}}=\sqrt{\frac{f(r_{ph})}{r_{ph}^2}} =\sqrt{V_{eff}(r_{ph})}
\end{eqnarray}
Thus, the bound on the magnitude of acceleration $|a|_b$ of LUA trajectories is equal to the square root of the maximum value of effective potential for the photon trajectories which gives the shadow radius of a black hole as given in Eq.(\ref{shadow radius}).

Collecting all the relations together, we can hence say that the parameters $r_{ph}$ and $r_{sh}$ defining the black hole shadow and the acceleration bounds $|a|_b$ and $r_b$ for the radial LUA trajectories in the Schwarzschild type black hole spacetime defined by the metric in Eq.(\ref{general class of spacetime}) are related as,
\begin{eqnarray}
r_{ph} &=& r_b \label{equivalence of photon sphere}\\
\text{and}\quad r_{sh} &=& \frac{1}{|a|_b}\label{equivalence of shadow}
\end{eqnarray}
This is indeed a fascinating correspondence since the notion of black hole shadow is completely independent of the LUA trajectories.

\section{For a general class of spherically symmetric black hole spacetime} \label{The correspondence in AB type BHs}

We further explore the connection between the shadow parameters and the acceleration bounds in a more general class of spherically symmetric black hole spacetimes. Consider a asymptotically flat black hole spacetime given by metric,
\begin{eqnarray}
{ds}^{2} &=& A(r) \, {dt}^{2}- B(r)\,{dr}^{2}-r^{2}\,d\Omega_{d-2}^{2} \label{AB class of spacetime}
\end{eqnarray}
where $A(r)$ and $B(r)$ are smooth differentiable functions with $A(r)\to 1$ and $B(r)\to 1$ as $r\to \infty$. The Killing horizon of the black hole is at $r=r_H$ and is obtained as a solution to $A(r_H)=0$ \cite{Johannsen}. The black hole solutions of this form have been found in many different gravity theories such as $f(R)$ gravity, Born-Infeld gravity, teleparallel gravity and Gauss-Bonnet gravity, see \cite{power-law Maxwell f(T) gravity,3D f(R) gravity, Born-Infeld gravity, Nontrivial BH in f(R) gravity, teleparallel gravity, ghost-free Gauss-Bonnet gravity} for examples.

By using the technique similar to the one used for the Schwarzschild type black hole, one can obtain the photon sphere radius and the shadow radius in terms of the metric functions $A(r)$ and $B(r)$, see \cite{Review} for a detailed derivation. As the metric is independent of $t$ and $\phi$, the energy $E$ and angular momentum $L$ of photons are constants of motion on null geodesics. The radial equation of motion for equatorial null geodesics in such a background spacetime is then given by,
\begin{eqnarray}
\frac{\dot{r}^2}{L^2}-\frac{1}{B(r)} \left(\frac{1}{b^2 A(r)}-\frac{1}{r^2}\right) &=& 0
\label{rconstraintnew}
\end{eqnarray}
where $b=L/E$ is the impact parameter. From the above equation, the effective potential can be read off as,
\begin{eqnarray}
V_{eff}(r,b) &=& -\frac{1}{B(r)} \left(\frac{1}{b^2 A(r)}-\frac{1}{r^2}\right)
\end{eqnarray}
As expected, the effective potential is a function of both the radial coordinate $r$ and the impact parameter $b$ and depends on both the metric functions $A(r)$ and $B(r)$. By definition of the photon sphere and the critical impact parameter as discussed in section \ref{The Shadow}, the constraints on the effective potential at $b=b_{cr}$ and $r=r_{ph}$ are,
\begin{eqnarray}
V_{eff}(r,b)\bigg|_{b=b_{cr},\, r=r_{ph}} = 0 \quad &\text{and}& \quad \frac{dV_{eff}(r,b)}{dr}\bigg|_{b=b_{cr},\, r=r_{ph}} = 0
\end{eqnarray}
To obtain the values of critical impact parameter $b_{cr}$ and the photon sphere radius $r_{ph}$ one needs to solve both the constraint equations simultaneously. One then gets the critical impact parameter in terms of the metric function $A(r)$ as,
\begin{eqnarray}
b_{cr} &=& \frac{2 \sqrt{A(r_{ph})}}{A'(r_{ph})} \label{impact parameter for AB}
\end{eqnarray}
where the prime denotes the derivative with respect to radial coordinate $r$ and the radius of photon sphere $r_{ph}$ can be obtained as a solution to the equation,
\begin{eqnarray}
r_{ph}\, A'(r_{ph}) - 2\, A(r_{ph}) &=& 0 \label{photon sphere for AB}
\end{eqnarray}
Using the radial equation Eq.(\ref{rconstraintnew}), we can get the radius of shadow seen by the observer at $r=r_0$ as
\begin{eqnarray}
 r_{sh} &=& r_0^2\frac{d\phi}{dr} \Bigg|_{r=r_0}  = \left[- V_{eff}(r_0,b_{cr})\right]^{-\frac{1}{2}}
\end{eqnarray}
Far away from the black hole as $r_0\to \infty$, the spacetime is flat and the effective potential reduces to $-1/b^2$. Thus far away from the black hole, we get the radius of black hole shadow $r_{sh}=b_{cr}$. The expression for shadow radius $r_{sh}$ can be further simplified using the condition on photon sphere in Eq.(\ref{photon sphere for AB}) to get,
\begin{eqnarray}
r_{sh} &=& \frac{r_{ph}}{\sqrt{A(r_{ph})}} \label{shadow radius for AB}
\end{eqnarray}
Altogether we have the shadow parameters namely, the photon sphere radius $r_{ph}$ given by Eq.(\ref{photon sphere for AB}) and the shadow radius far away from the black hole given by Eq.(\ref{shadow radius for AB}). One can note that both the shadow parameters depend only on the metric function $A(r)$ i.e the $g_{tt}$ component of the metric.

Now, to compare the shadow parameters to acceleration bounds we need to obtain the solutions for the radial LUA trajectories in this general static spherically symmetric background spacetime. It turns out that one can setup the equations for the linearity conditions $w^i-|a|^2 u^i=0$ in this case; but the equations are not analytically solvable to get a general solution in terms of the metric functions $A(r)$ and $B(r)$ as in the case of Schwarzschild type black holes. Hence we choose a simpler, analytically tractable special case within this category of black hole spacetimes wherein the metric components $A(r)$ and $B(r)$ are related through a constant, say $c$, such that $B(r)= c\, {A(r)}^{-1}$ with $c>0$. For this special case, the solution for LUA trajectory can be obtained analytically in terms of the proper-velocity components as in the case of Schwarzschild type black holes. In the following section, we investigate the acceleration bounds of the LUA trajectories and their relation to the shadow parameters in this special class of black hole spacetimes.

\subsection{Special case of static spherically symmetric spacetime} \label{special}

Considering the relation between the metric functions $A(r)$ and $B(r)$ as $B(r)= c\, {A(r)}^{-1}$, the metric of the static spherically symmetric black hole becomes,
\begin{eqnarray}
{ds}^{2} &=& A(r) \, {dt}^{2}-c {A(r)}^{-1}{dr}^{2}-r^{2}\,d\Omega_{d-2}^{2} \label{AcA class of spacetime}
\end{eqnarray}
where now we have $A(r_H)=0$ at the horizon $r_H$ of the black hole. One such black hole solution in bumblebee gravity is given in \cite{bumblebee gravity}.

As both the photon sphere radius and the shadow radius are function of only the $g_{tt}$ component of the metric, the expressions in Eq.(\ref{photon sphere for AB}) and (\ref{shadow radius for AB}) are still valid in the above black hole spacetime.

The differential equations obtained as the linearity conditions $w^i-|a|^2 u^i=0$ with the constant magnitude of acceleration $|a|$ are solvable analytically and the solution to the radial LUA trajectory in this background spacetime can be obtained in terms of its proper-velocity components. The non-zero components of the velocity vector of a LUA trajectory are obtained to be,
\begin{eqnarray}
\frac{dt}{d\tau}&=&\sqrt{c}\,{A(r)}^{-1}\left(|a|r+h\right)\label{AcA time velocity}\\
\frac{dr}{d\tau}&=&\pm\sqrt{\left(|a|r+h\right)^2-\frac{A(r)}{c}}\label{AcA radial velocity}
\end{eqnarray}
Now, following the procedure as used for obtaining the LUA trajectory in the spacetime of Schwarzschild type black holes in section \ref{The correspondence in Sch type BHs}, we proceed to obtain the bound on acceleration $|a|_b$ and the bound on distance of closest approach $r_b$ in the black hole spacetime given by metric in Eq.(\ref{AcA class of spacetime}). We start by writing the radial component of solution with $h=0$ as,
\begin{eqnarray}
\left(\frac{d r}{d\tau}\right)^2  &=&  |a|^2r^2 - \frac{A(r)}{c}
\end{eqnarray}
The effective potential for the radial motion of LUA trajectory can be read off as,
\begin{eqnarray}
\hat{V}_{\mathit{eff}}(r) &=& -\left[|a|^2r^2 - \frac{A(r)}{c}\right]
\end{eqnarray}
The bounds on acceleration $|a|_b$ and on turning point $r_b$ can be arrived at using the constraints on the effective potential as given in Eqs.(\ref{constrain 1 on effective potential}) and (\ref{constrain 2 on effective potential}). For the special metric under consideration, these conditions lead to the following equations,
\begin{eqnarray}
c\, |a|_b^2 \, r_{b}^2 - A(r_{b}) &=& 0 \label{AcA potential=0}\\
2\, c |a|_b^2 \, r_b - A'(r_b) &=& 0 \label{AcA derivative of potential=0}
\end{eqnarray}
Solving the above equations simultaneously, we get the acceleration bound $|a|_b$ to be,
\begin{eqnarray}
|a|_b &=& \frac{1}{\sqrt{c}} \frac{\sqrt{A(r_b)}}{r_b} \label{AcA acceleration bound}
\end{eqnarray}
whereas the bound on the distance of closest approach $r_b$ can be obtained through,
\begin{eqnarray}
2 A(r_b) -r_b A'(r_b) &=& 0 \label{AcA constraint equation for the bound on turning point}
\end{eqnarray}
One can note that the above equation is same as that for the radius of photon sphere in Eq.(\ref{photon sphere for AB}) which implies that the bound on the turning point $r_b$ will always be equal to the radius photon sphere $r_{ph}$ in all static spherically symmetric black hole spacetimes of the form of Eq.(\ref{AcA class of spacetime}). Using Eq.(\ref{AcA constraint equation for the bound on turning point}) in Eq.(\ref{AcA acceleration bound}), one can get a relation between the bound on acceleration $|a|_b$ amd the shadow radius given in Eq.(\ref{shadow radius for AB}) as,
\begin{eqnarray}
|a|_b &=& \frac{1}{\sqrt{c}}\frac{\sqrt{A(r_b)}}{r_b}=\frac{1}{\sqrt{c}}\frac{\sqrt{A(r_{ph})}}{r_{ph}} =\frac{1}{\sqrt{c}}\, \frac{1}{r_{sh}}
\end{eqnarray}
Collecting the results, we can summarize the connection between the shadow parameters $r_{ph}$ and $r_{sh}$ and the acceleration bounds $|a|_b$ and $r_b$ of the LUA trajectories in the static spherically symmetric black hole spacetime defined by metric in Eq.(\ref{AcA class of spacetime}) as,
\begin{eqnarray}
r_{ph} &=& r_b \label{AcA equivalence of photon sphere}\\
r_{sh} &=& \frac{1}{\sqrt{c}}\frac{1}{|a|_b}\label{AcA equivalence of shadow}
\end{eqnarray}
Here the photon sphere radius is equal to the bound on the distance of closest approach as obtained in the case of Schwarzschild type black hole spacetimes whereas the shadow radius doesn't satisfy the same relation as in Eq.(\ref{equivalence of shadow}). However, the shadow radius $r_{sh}$ can be made equal to the inverse of acceleration bound $|a|_b$ for a different value of the boundary data $h\neq 0$ while compromising the equality between the photon sphere radius $r_{ph}$ and the bound on distance of closest approach $r_b$. This we have checked numerically for some parameters ranges for a specific black hole solution. Hence the relations in Eqs.(\ref{equivalence of photon sphere}) and (\ref{equivalence of shadow}) are valid simultaneously only in the special case of Schwarzschild type black hole spacetimes for boundary data $h=0$. In a more general black hole spacetimes of the form of Eq.(\ref{AcA class of spacetime}), the two connections are valid separately for different boundary data values $h$. In the most general static spherically symmetric black hole with the metric defined by Eq.(\ref{AB class of spacetime}), it is not possible to comment whether such connections will hold in these background spacetimes or not, since the LUA trajectory is not analytically solvable.

\section{Discussion}\label{discussion}

In this work, we explored an interesting connection between the shadow parameters of black holes in static spherically symmetric geometries of the Schwarzschild type as defined by the metric in Eq.(\ref{general metric}) and the acceleration bounds for radial linear uniformly accelerated trajectories, which are the generalizations of the Rindler trajectories, in the background of the same black hole geometries. We found that for a particular choice of boundary data $h=0$ for LUA trajectory, the photon sphere radius $r_{ph}$ is equal to the lower bound on radius of closest approach $r_b$ of the radial LUA trajectory incoming from past null infinity while the shadow radius $r_{sh}$ of the image of black hole created due to uniform background illumination as seen by an observer far away is equal to the inverse of magnitude of the acceleration bound $|a|_b$ for the LUA trajectory to turn back to infinity. The relations are simply, 
\begin{equation}
r_{sh}=\frac{1}{|a|_b} \quad\quad \text{and} \quad\quad r_{ph}= r_b. \label{Final result}
\end{equation}
We first checked the relation by explicitly solving and obtaining the expressions for these parameters in the case of a $d$-dimensional charged black hole in Einstein's theory of gravity, a vacuum black hole solution in the pure Lovelock theory of gravity and a black hole with conformally coupled scalar field in the $\mathcal{F}(R)$ theory of gravity. We then showed, using the effective potential technique, that the above relations hold for a general form of the metric function $f(r)$ for the static spherically symmetric geometries of the Schwarzschild type defined by Eq.(\ref{general metric}) and thus would hold in any theory of gravity with a metric solution of the similar form.

We further investigated the connection beyond the Schwarzschild type black holes, for a special case within the general class of static spherically symmetric spacetimes wherein the metric components $g_{tt}$ and $g_{rr}$ are represented by two different functions of radial coefficient $r$, say $A(r)$ and $-B(r)$ respectively. In the most general static spherically symmetric black hole spacetime the LUA trajectory is not analytically solvable; whereas the shadow parameters $r_{ph}$ and $r_{sh}$ can be evaluated explicitly. In the special case where the black hole metric components are related through $B(r)=c/A(r)$ with a positive constant $c$, we found an analytical solution for the radial LUA trajectory and further obtained the relations between the shadow parameters and the acceleration bounds for the boundary data $h=0$. In this case, the photon sphere radius $r_{ph}$ and the bound on the distance of closest approach $r_b$ are found to be equal as in Eq.(\ref{Final result}) whereas the shadow radius $r_{sh}$ and the bound $|a|_b$ are now related through the constant $c$. It thus seems that the direct and simultaneous connection between $\{r_{ph} , r_{sh}\}$ and $\{r_{b} , |a|_{b}\}$ exists only for the Schwarzschild type black holes whereas for the special case discussed in section \ref{special} the connection works for $\{r_{ph}\}$ and $\{r_{b}\}$ or $\{r_{sh}\}$ and $\{ |a|_{b}\}$ separately.

For the  Schwarzschild type black holes, the relations in Eq.(\ref{Final result}), are indeed intriguing since kinematics of the photon sphere and the shadow of a black hole involve null geodesics while the kinematics of the acceleration bounds involve LUA trajectories and both seem independent of each other. However, they both involve limiting cases in their own respective domains. Perhaps the relations are just an algebraic coincidence or perhaps they signify a deeper connection between the two aspects; the black hole gravitational potential being the common denominator in the  Schwarzschild type black holes; which needs to be investigated further.

\end{document}